\documentstyle[preprint,aps,epsf]{revtex}

\newcommand{\postscript}[2]{\setlength{\epsfxsize}{#2\hsize}
 \centerline{\epsfbox{#1}}}
\newcommand{\oln}{\overline}

\begin{document}

\tightenlines

\title{Universal description of $S$-wave meson spectra in a renormalized
       light-cone QCD-inspired model}

\author{T. Frederico}
\address{Dep.de F\'\i sica,
         Instituto Tecnol\'ogico de Aeron\'autica,
         Centro T\'ecnico Aeroespacial, \\
         12.228-900 S\~ao Jos\'e dos Campos, S\~ao Paulo, Brazil}

\author{Hans-Christian Pauli}
\address{Max-Planck Institut f\"ur Kernphysik, D-69029 Heidelberg, Germany}

\author{Shan-Gui Zhou}
\address{Max-Planck Institut f\"ur Kernphysik, D-69029 Heidelberg, Germany \\
         and
         School of Physics, Peking University, Beijing 100871, China}

\date{Draft \today}

\maketitle

\begin{abstract}
A light-cone QCD-inspired model, with the mass squared operator
consisting of a harmonic oscillator potential as confinement and a
Dirac-delta interaction, is used to study the $S$-wave meson
spectra. The two parameters of the harmonic potential and quark
masses are fixed by masses of $\rho(770)$, $\rho(1450)$, $J/\psi$,
$\psi(2S)$, $K^*(892)$ and $B^*$. We apply a renormalization
method to define the model, in which the pseudo-scalar ground
state mass fixes the renormalized strength of the Dirac-delta
interaction. The model presents an universal and satisfactory
description of both singlet and triplet states of $S$-wave mesons
and the corresponding radial excitations.
\\
PACS numbers: 11.10.Ef, 11.10.Gh, 12.39.Hg, 12.39.Ki, 14.40-n
\end{abstract}


\section{Introduction}
\label{sec:intro}

In the effective light-cone QCD theory~\cite{BRO98,PAU00} the
lowest Fock component of the hadron wave function is an
eigenfunction of an effective mass squared operator with
constituent quark degrees of freedom and parameterized in terms of
an interaction which contains a Coulomb-like potential and a
Dirac-delta term. The Fock-state components of the hadron
light-front wave function can be constructed recursively from the
lowest Fock-state component. The interaction in the mass operator
comes from an effective one-gluon-exchange where the Dirac-delta
term corresponds to the hyperfine interaction. The masses of the
ground state of the pseudo-scalar mesons and in particular the
pion structure~\cite{FRE01} were described reasonably, with a
small number of free parameters, which is only the canonical
number plus one --- the renormalized strength of the Dirac-delta
interaction.

The model was extended to include the confining interaction and
used to study the splitting of the excited pseudo-scalar states
from the excited $^3S_1$ vector meson states as a function of the
ground state pseudo-scalar mass~\cite{FRE02}. In
Ref.~\cite{FRE02}, the Coulomb-like and the confining interactions
were substituted by a harmonic oscillator potential, which allowed
an analytic formulation. The parameters of the confining
interaction in the mass squared operator, were fitted to the
$^3S_1$-meson ground state mass and to the slope of the trajectory
of excited states with the radial quantum number~\cite{ANI00}.
With the renormalized strength of the Dirac-delta interaction
fixed by pseudo-scalar masses, it was shown that the $\pi$-$\rho$
mass splitting, due to the attractive Dirac-delta interaction, is
the source of the splitting between the masses of the excited
states. A reasonable agreement with the data~\cite{PDG02} was
obtained. Besides, in light-cone framework (mass squared operator
appears in the Hamiltonian), the Dirac-delta plus confining
harmonic oscillator potential gave a natural explanation of the
observation of the almost linear relationship between the mass
squared of excited states with radial quantum number
$n$~\cite{ANI00}. This reveals some of the physics that are
brought by the work of Ref.~\cite{ANI00} and showed the relation
between the $\pi$ and $\rho$ spectrum, through the pion mass scale
which defines the renormalization condition of the model.

In the present paper, this simple model, with the Dirac-delta
interaction acting in the $^1S_0$ channel only and harmonic
oscillator potential as confinement, is used to investigate the
$S$-wave meson spectra from $\pi$-$\rho$ to $\eta_b$-$\Upsilon$
and make predictions for $\eta_t$-$\theta$ (we do not study
$\eta$-$\omega$ and $\eta'$-$\phi$). Instead of using
flavor-dependent parameters, parameters are found for the harmonic
oscillator potential which is valid for not only light mesons but
also heavy mesons. In other words, the parameters are universal.
It is shown that the linear relationship between the mass squared
of excited states with radial quantum number is still
qualitatively valid even for heavy mesons like $\Upsilon$. The
simply model presents reasonable agreement with available data
and/or with the meson mass spectra given by Godfrey and
Isgur~\cite{GOD85}.

This paper is organized as follows. In section~\ref{sec:theory},
we give very briefly the extension of the light-cone QCD-inspired
theory for which the mass squared operator of a constituent
quark-antiquark system includes a confining
interaction~\cite{FRE02}. The renormalization of the theory using
the subtracted equations for the transition matrix~\cite{FRE00} of the model
can be found in Ref.~\cite{FRE01,FRE02} thus is omitted here. In
the same section, we present the Dirac-delta term plus harmonic
oscillator potential approach and solve it with the $T$ matrix
method developed in Ref.~\cite{FRE01,FRE02}. The results and
discussion are presented in section~\ref{sec:results}. A brief
summary is given in the last section.

\section{Extended Light-cone QCD-inspired theory with
         Dirac-delta and harmonic oscillator confining potential}
\label{sec:theory}

In this section we review our previous work~\cite{FRE02}, in which
we have extended the renormalized effective QCD-theory of
Ref.~\cite{FRE01} to include confinement. In the effective theory
the bare mass operator equation for the lowest Light-Front
Fock-state component of a bound system of a constituent quark and
antiquark of masses $m_1$ and $m_2$, is described
as~\cite{BRO98,PAU00}
\begin{eqnarray}
 M^2 \psi (x,{\vec k_\perp})
 & = &
  \left[ \frac{{\vec k_\perp}^2+m^2_1}{x}
       + \frac{{\vec k_\perp}^2+m^2_2}{1-x}
  \right]
  \psi (x,{\vec k_\perp})
\nonumber \\
 &   & \mbox{}
  - \int \frac{dx' d{\vec k'_\perp}\theta(x')\theta (1-x')}
              {\sqrt{x(1-x)x'(1-x')}}
  \left( \frac{4m_1m_2}{3\pi^2} \frac{\alpha}{Q^2}
       - \lambda - W_{\text{conf}}(Q^2)
  \right)
  \psi (x',{\vec k'_\perp}) \ ,
\label{p1}
\end{eqnarray}
where $M$ is the mass of the bound-state and $\psi$ is the
projection of the light-front wave-function in the quark-antiquark
Fock-state. The confining interaction is included in the model by
$W_{\text{conf}}(Q^2)$. The momentum transfer $Q$ is the
space-part of the four momentum transfer and the strength of the
Coulomb-like potential is $\alpha$. The singular interaction is
active only in the pseudo-scalar meson channel with $\lambda$ as
the bare coupling constant.

For convenience the mass operator equation is transformed to the
instant form representation~\cite{PAU00b}, which in operatorial
form is written as~\cite{FRE02}
\begin{eqnarray}
 \left( M_0^2 + V + V^\delta +V_{\text{conf}} \right) | \varphi \rangle
 = M^2 |\varphi \rangle
 \ ,
 \label{mass2}
\end{eqnarray}
where the free mass operator, $M_0 \ (=E_1+E_2) $, is the sum of
the energies of quark 1 and 2 ($E_i=\sqrt{m_i^2+k^2}$, $i$=1, 2
and $k\equiv|\vec k|$), $V$  is the Coulomb-like potential,
$V^\delta$ is the short-range singular interaction and
$V_{\text{conf}}$ gives the quark confinement.

We simplify Eq.~({\ref{mass2}) by omitting the Coulomb term to the
form~\cite{FRE02}
\begin{eqnarray}
 \left( M^2_{\rm ho} + g \delta (\vec r) \right) \varphi (\vec r)
 = M^2 \varphi (\vec r)
 \ ,
 \label{mass3}
\end{eqnarray}
where the bare strength of the Dirac-delta interaction is $g$, and
the mass squared operator is~\cite{PAU00b}
\begin{eqnarray}
 M^2_{\rm ho}= \left( C(k)k^2+ m_s^2 \right) + 2 m_s v(r)
 \ ,
 \label{mass4}
\end{eqnarray}
in units of $\hbar = c =1$, $m_s = m_1+m_2$. The dimensionless
factor of $k$ is
\begin{equation}
 C(k) = 2 + \frac{E_1+m_1}{E_2+m_2} + \frac{E_2+m_2}{E_1+m_1}.
\end{equation}
In the following, we approximate $C(k)$ as
${m_s}/{m_r}$~\cite{PAU00b}, with $m_r=m_1m_2/(m_1+m_2)$.

The harmonic oscillator potential is introduced as a confinement
\begin{equation}
 v(r) = - c_0 + \frac{1}{2} c_2 r^2
 \ ,
\end{equation}
where $c_0$ and $c_2$ are two universal parameters valid for all
of the mesons. The eigenvalue Eq.~(\ref{mass4}) is given now by
\begin{equation}
  \displaystyle
  2m_s \left(-\frac{1}{2m_r}\nabla^2 + \frac{1}{2} c_2 r^2
    + \frac{1}{2} m_s - c_0 \right) \Psi_{n}(\vec r)
    = M^2_{n} \Psi_{n}(\vec r),
 \label{mass5}
\end{equation}
with $\Psi_{n}(\vec r)$ the eigenstate of the harmonic oscillator
potential and the corresponding eigenvalue
\begin{equation}
M^2_{n} = 2m_s \left( \left( 2n+\frac{3}{2}\right) \sqrt{c_2/m_r}
                      + \frac{1}{2} m_s - c_0 \right)
         = n w + m^2_s + 3m_s\sqrt{c_2/m_r} - 2m_s c_0,
 \label{eq:masssquare}
\end{equation}
where $n$ (0, 1, 2, $\cdots$) is the radial quantum number and
$w=4m_s\sqrt{c_2/m_r}$. Note that here $n$ begins from 0 for
convenience, while in discussions in Section~\ref{sec:results}, it
begins from 1 to keep accordance with literature. In the present
model, Eq.~(\ref{eq:masssquare}) gives the vector meson spectrum
since Dirac-delta interaction acts only on pseudo-scalar mesons.
The mass squared of the ground state ($n=0$) of the $^3S_1$ meson
can be written as
\begin{equation}
 M^2_{\rm gs}
 = m^2_s + 3m_s\sqrt{c_2/m_r} - 2m_s c_0 .
 \label{eq:gsmass}
\end{equation}

With the subtraction point, $\mu$, taken to be the mass of the
pseudo-scalar meson ground state, the reduced $T$ matrix was
derived in Ref.~\cite{FRE02} as
\begin{eqnarray}
 t^{-1}_{\cal R}(M^2) = (2\pi)^3\sum_{n} |\Psi_n(0)|^2
  \left(\frac{1}{\mu^2-M^2_{n}}-
        \frac{1}{M^2-M^2_{n}} \right)
 \ .
 \label{trenv23}
\end{eqnarray}
The value of the $S$-wave eigenfunction at the origin is given
by~\cite{GOL63}
\begin{eqnarray}
 \Psi_n(0) = \alpha^{\frac32}\left[ \frac{2^{2-n}}{\sqrt{\pi}}
              \frac{(2 n+1)!!}{n!}\right]^\frac12
 \ ,
 \label{wf0}
\end{eqnarray}
where $\alpha^{-1}$ is the oscillator length. The final form the
reduced $T$ matrix is
\begin{eqnarray}
 t^{-1}_{\cal R}(M^2)
 = (2\pi\alpha)^3\sum_{n=0}^\infty \frac{2^{2-n}}{\sqrt{\pi}}
    \frac{(2 n+1)!!}{n!}
    \left(\frac{1}{\mu^2- n w - M^2_{\text gs}}
         -\frac{1}{M^2- n w - M^2_{\text gs}}
    \right)
 \ .
 \label{trenv24}
\end{eqnarray}
The zeros of Eq.~(\ref{trenv24}) gives the eigenvalues of the the
mass squared operator of Eq.~(\ref{mass3}). In Ref.~\cite{FRE02},
$\mu$ was changed continuously to study the splitting between the
$^1S_0$ and $^3S_1$ spectrum of the $\pi$-$\rho$ mesons. In the
following section, we are going to discuss this splitting in more
detail and apply the model to mesons with heavy quarks as well.

\section{Results and discussions}
\label{sec:results}
\subsection{Parameters and nomenclature}
\label{subsec:parameters}

The masses of $\rho(770)$, $\rho(1450)$, $J/\psi(1S)$ and
$\psi(2S)$~\cite{PDG02} and Eqs.~(\ref{eq:masssquare},
\ref{eq:gsmass}) are used to fix $c_0$, $c_2$ and up, down and
charm quark masses with the assumption of $m_u=m_d$. Then masses
of strange and bottom quarks are determined by the masses of $K^*$
and $B^*$~\cite{PDG02} and Eq.~(\ref{eq:gsmass}). In order to
predict $t$-quark meson spectrum, we adopt an estimate from
Godfrey and Isgur~\cite{GOD85}, $m_t=35$ GeV. The parameters used
in the present model are listed in Table~\ref{table:parameters}.

The physical nomenclature of mesons are given in
Table~\ref{table:meson-nomenclature} to facilitate the following
discussion. Since we assume that up and down quarks are the same,
and since mixing between different flavors can not be dealt with
within this simple model, we investigate only $I=1$ states among
the diagonal meson sectors containing $u$, $d$ and $s$ quark.
$m_u=m_d$ also means that in our model the spectrum of $\pi^0$
($\rho^{0}$, $K^0$, $K^{0*}$, $D^0$, $D^{0*}$, $B^0$, $B^{0*}$,
$T^0$, $T^{0*}$) is same as that of $\pi^\pm$ ($\rho^\pm$,
$K^\pm$, $K^{\pm*}$, $D^\pm$, $D^{\pm*}$, $B^\pm$, $B^{\pm*}$,
$T^\pm$ and $T^{\pm*}$). In the following discussions, we will
omit the charge signs of mesons for simplicity.

\subsection{$S$-wave meson spectra}
\label{subsec:spectra}

The splitting of the light mesons, $\pi$-$\rho$ and $K$-$K^*$
spectra, due to the Dirac-delta interaction acting in the
pseudo-scalar $^1S_0$ channel, are studied in the previous
work~\cite{FRE02}. There the empirical slope ($w$) from
Ref.~\cite{ANI00} or Ref.~\cite{PDG02} are used directly. In the
present paper, $\pi$-$\rho$ and $K$-$K^*$ are reinvestigated in
the same framework as other $S$-state mesons. The results are
presented in Fig.~\ref{figure:pi} and Fig.~\ref{figure:k}
respectively. Similar agreement with the data as that in
Ref.~\cite{FRE02} is found in Fig.~\ref{figure:pi}, simply because
the present model gives $w=1.55$ GeV$^2$ which is very close to
$w=1.39$ GeV$^2$ used in Ref.~\cite{FRE02}. For $K^*$, the present
model gives a large value of $w$, 1.92 GeV$^2$, (due to the fact
that strange quark mass is larger than up-down quark mass, $w$ for
$K^*$ must be larger than that for $\rho$ from the present model),
compared to that extracted from the data, 1.19 GeV$^2$. However,
as pointed out in Refs.~\cite{ANI00,PDG02}, there are ambiguities
about the quantum number assignment of exited states of $K$ and
$K^*$. If we follow the identification of the quark
model~\cite{PDG02} for $K(1460)$ ($2^1S_0$), $K(1830)$ ($3^1S_0$)
and $K^*(1410)$ ($2^3S_1$), the spectra of $K$-$K^*$ from present
model are in good agreement with the data.

In Fig.~\ref{figure:etac} the results are shown for the
$\eta_c$-$\psi$ mass splitting as a function of the ground-state
pseudo-scalar mass $\mu$, which interpolates from the $\eta_c$ to
the $J/\psi$ meson spectrum. Compared to the light $\pi$-$\rho$
and $K$-$K^*$ mesons, the splitting is smaller even for the ground
state ($\sim$ 100 MeV) and becomes weaker in the excited states,
although the model attributes consistently smaller masses for the
$^1S_0$ states compared to the respective $^3S_1$ ones. An excited
state of $\eta_c$ is observed without definite spin and parity
assignment~\cite{PDG02}. From our model, this state might be
$\eta_c(2S)$, which is consistent with the assignment of quark
model~\cite{PDG02}. There are many exited states for $\psi$, such
as $\psi(3770)$, $\psi(4040)$, $\psi(4160)$, and $\psi(4415)$, all
of them are assigned to $J^\pi=1^-$. Considering that $\psi(3770)$
is $\psi(1^3D_1)$~\cite{PDG02,GOD85}, $\psi(4040)$ or $\psi(4160)$
seems to be $\psi(3^3S_1)$ from the present model.

From Eqs.~(\ref{eq:masssquare}, \ref{eq:gsmass}) and quark masses
listed in Table~\ref{table:parameters}, one obtains the spectrum
of $D^{*}$. The splitting of $D$-$D^{*}$ and spectrum of $D$ meson
can be calculated from Eq.~(\ref{trenv23}) with the renormalized
strength of $\delta$ potential fixed by the mass of corresponding
pseudo-scalar ground state $D$ and is shown in
Fig.~\ref{figure:D}. One finds good agreement for $D^{*}(1S)$ with
the data. The predicted mass of $D(2S)$ is about 10\% larger than
the unconfirmed data.

Similarly, the present model gives $D^*_s(1S)$ with good agreement
with the data as is seen from Fig.~\ref{figure:Ds} in which the
spectra for $D_s$-$D^*_s$ are presented.

As mentioned before, the bottom quark mass is fixed by mass of
$B^*(1S)$ and Eq.~(\ref{eq:gsmass}). There are no more data for
$b$-quark mesons. The predicted spectra for $B$ and $B^*$ are
presented in Fig.~\ref{figure:B}.

The spectrum of $\Upsilon$ can be calculated from our model with
parameters listed in Table~\ref{table:parameters} and is shown in
Fig.~\ref{figure:etab}. We should note that the agreement with the
data is good, from $\Upsilon(1S)$ to $\Upsilon(4S)$, considering
that no parameter is adjusted specially for $\Upsilon$, the
heaviest meson states observed up to now. The unconfirmed
bottomonium state $\eta_b(9300)$ is used to predict the spectrum
of $\eta_b$ and also presented in the same figure.

No confirmed data are available for the triplet states of $B_s$
and $B_c$. With the bottom quark mass $m_b=5068$ MeV, the present
model predicts the triplet ground state masses for $B_s$ and
$B_c$, being 5342.1 MeV and 6.346 GeV, respectively. Although the
predicted mass for $B_s(1S)$ is comparable to the unconfirmed data
5416.6 MeV, both predictions are respectively smaller than their
corresponding singlet ground state masses, 5369.6 MeV and 6.4 GeV.
Therefore we can not calculate within our model the singlet
spectra $B_s$ and $B_c$. However, for completeness and in order to
give a reference to the reader, we list the calculated values for
triplets of the bottom strange and bottom charged mesons in
Table~\ref{table:Bs} and Table~\ref{table:Bc} respectively.

The vector top meson spectra from the present model with the top
quark mass $m_t$ = 35 GeV~\cite{GOD85} and $m_t$ = 175
GeV~\cite{PDG02} are given in Table~\ref{table:top}. The top quark
mass $m_t$ = 35 GeV is used in order to compare the present
spectra with predictions of Ref.~\cite{GOD85}. Qualitative
agreement between the two models is found.

\subsection{Further discussions}
\label{subsec:discussion}

From the above results, it is clear that our model, with the mass
squared operator consisting of a harmonic oscillator confining
potential and a Dirac-delta interaction, could be used to describe
universally and satisfactorily both singlet and triplet states of
$S$-wave mesons as well as radial excitations. However, lattice
QCD calculations predict that the quark-antiquark potential
increases linearly with the distance between quark and antiquark
$r$ when $r$ is large. One may ask, does this contradict our
model? The answer is no. In the following, we justify roughly that
the two interactions are consistent with each other for large $r$
disregarding the case for small $r$ which is not important in the
present work.

In the front form of QCD, the mass squared $M^2 = m_s^2 + 2m_s E =
m_s^2 + 2m_s (T_{\text {FF}}+V_{\text {FF}})$ ($m_s$ is the sum of
the quark and antiquark mass, $T$ is the kinetic
energy and $V_{\text {FF}}$ is the interaction)~\cite{PAU00b},
while in the instant form of QCD, $M = m_s
+ E = m_s + (T_{\text {IF}}+V_{\text {IF}}$). The relation between
the interactions $V_{\text {FF}}$ and $V_{\text {IF}}$ is found for large $r$
(where $T\ll V\ll V^2$) as $V_{\text {FF}} \sim V^2_{\text {IF}}$,
from which the harmonic like potential for the mass squared
operator in the front form can be derived
from  the linear confinement potential in the instant
form of QCD.

Very recently the radial excitations of light mesons were studied
in detail in the framework of the QCD string
approach~\cite{BAD02}. There the spin-averaged meson masses were
calculated with a modified confining potential and the calculated
slopes of the radial Regge trajectories are in agreement with
Ref.~\cite{ANI00}. In Ref.~\cite{BAD02}, the linear relation
between mass squared $M^2_n$ and the radial quantum number $n$
comes mainly from properties of the approximated eigenvalues of
the spinless Salpeter equation with the linear confining
potential. Since in our light-cone QCD model, the {\em harmonic
oscillator potential} is included in the {\em mass squared}
operator, one arrives naturally at the same linear relation
between $M^2_n$ and $n$ for vector mesons. This gives an explicit
explanation of the radial Regge trajectories found for light
vector mesons in Ref.~\cite{ANI00}. This relation is still valid
even for heavy vector mesons as shown in the previous subsection. In
addition, an extension of the present model with orbital
excitation included could also be used to describe the orbital
Regge trajectory.

For pseudoscalar mesons, particularly for the light ones, the present
model does not support the simple linear relation because the
Dirac-delta interaction plays an important role now. However, this
is not in contradiction with the data because light pseudoscalar
mesons follow the radial Regge trajectories poorly. Let's take the
pion as an example, the slope $w$ = 1.67 GeV$^2$ is derived from
the masses of $\pi$ ($M$ = 0.14 GeV) and its first radial excited
state ($M$ = 1.3 GeV), while a slope smaller by $\sim$ 10\%, $w =
1.55$ GeV$^2$ can be calculated from its first and second excited
states ($M = 1.8$ GeV).

\section{Summary}
\label{sec:summary}

We applied the renormalized light-cone QCD-inspired effective
theory with confinement in the mass squared operator to study
mesons with heavy quarks. For the confinement, the harmonic
oscillator potential is used which allows an analytic solution of
our model. The Coulomb-like potential is omitted in the present
work while the Dirac-delta interaction is kept which acts on the
singlet $S$-wave states and plays important role in the splitting
of the singlet-triplet spectra. The two parameters of the harmonic
potential and quark masses are fixed by masses of $\rho(770)$,
$\rho(1450)$, $J/\psi$, $\psi(2S)$, $K^*(892)$ and $B^*$. A $T$
matrix renormalization method is used to renormalize the model, in
which the pseudo-scalar ground state mass fixes the renormalized
strength of the Dirac-delta interaction.

The model is applied to study the $S$-wave meson spectra from
$\pi$-$\rho$ to $\eta_b$-$\Upsilon$ and is also used to predict
top quark meson spectra. The linear relationship between the mass
squared of excited states with radial quantum
number~\cite{ANI00}---the radial Regge
trajectory~\cite{BAD02}---is apparent from our model and is found
to be qualitatively valid even for heavy mesons like $\Upsilon$.

The simple model presents satisfactory agreement with available
data and/or with the meson mass spectra given by Godfrey and
Isgur~\cite{GOD85}. Therefore, the recently proposed extension of
the light-cone QCD-inspired model which includes confinement while
keeping simplicity and renormalizability, gives a reasonable
picture of the spectrum of both light and heavy mesons. An
extension of the present model with orbital excitations included
could also be used to describe the orbital Regge trajectories.

\acknowledgments

TF thanks the Max Planck Institute of Heidelberg and specially to
Prof. Hans-Christian Pauli for the warm hospitality during the
period where this work was initiated. TF also thanks Conselho
Nacional de Desenvolvimento Cient\'\i fico e Tecnol\'ogico (CNPq)
and Funda\c c\~ao de Amparo a Pesquisa do Estado de S\~ao Paulo
(FAPESP) of Brazil for partial financial support. SGZ is partly
supported by the Major State Basic Research Development Program of
China Under Contract Number G2000077407 and by the Center of
Theoretical Nuclear Physics, National Laboratory of Heavy Ion
Accelerator at Lanzhou, China.

\begin{table}
\caption{
Parameters used in the present paper. $c_0$ and $c_2$ of harmonic
oscillator potential and masses of up, down and charm quarks are
fixed from masses of $\rho(770)$, $\rho(1450)$, $J/\psi(1S)$ and
$\psi(2S)$, with the assumption of $m_u=m_d$. Strange and bottom
quark masses are determined by masses of $K^*$ and $B^*$. The top
quark masses, $m_t$ = 35 GeV, same as that in Godfrey et
al.~\protect\cite{GOD85}, and $m_t$ = 175 GeV
from~\protect\cite{PDG02}, are used to predict spectra for
$t$-quark mesons. The data for meson masses are taken from
Hagiwara et al.~\protect \cite{PDG02}.
}
\label{table:parameters}
\begin{tabular}{lccccccc}
 Parameters & $c_0$ [MeV] & $c_2$ [GeV$^3$] & $m_u=m_d$ [MeV] & $m_s$ [MeV] &
 $m_c$ [MeV] & $m_b$ [MeV] & $m_t$ [GeV]\\
 \hline
 Values     & $807$       & 7.13$\times 10^{-2}$ & 265        & 478         &
 1749        & 5068       & 35/175      \\
\end{tabular}
\end{table}

\begin{table}
\newcommand{\fpirho}{\frame{$\pi^0,\eta,\eta'||\rho^0,\omega,\phi$}}
\newcommand{\etaeta}{\frame{$\eta,\eta'||\omega,\phi$}}
\newcommand{\fetacp}{\frame{$\eta_c||\psi$}}
\newcommand{\fetabU}{\frame{$\eta_b||\Upsilon$}}
\caption{
The physical nomenclature of the mesons as a reminder. Pseudo-scalar
mesons are given on the left, vector mesons on the right of each
sector. Diagonal sectors are marked with frames for guiding eye.
}
\label{table:meson-nomenclature}
\begin{tabular}{c|cccccc}
     & $\oln d$                & $\oln u$                & $\oln s$            &
       $\oln c$                & $\oln b$          & $\oln t$                    \\
 \hline
 $d$ & \fpirho                 & $\pi^-||\rho^-$ & $K^0||K^{*0}$               &
       $D^-||D^{*-}$           & $B^0||B^{*0}$     & $T^-||T^{*-}$               \\
 $u$ & $\pi^+||\rho^+$         & \fpirho         & $K^+||K^{*+}$               &
       $\oln D^0||\oln D^{*0}$ & $B^+||B^{*+}$     & $\oln T^0||\oln T^{*0}$     \\
 $s$ & $\oln K^0||\oln K^{*0}$ & $K^-||K^{*-}$   & \etaeta                     &
       $D_s^-||D_s^{*-}$       & $B_s^0||B_s^{*0}$ & $T_s^-||T_s^{*-}$           \\
 $c$ & $D^+||D^{*+}$           & $D^0||D^{*0}$   & $D_s^+||D_s^{*+}$           &
       \fetacp                 & $B_c^+||B_c^{*+}$ & $\oln T_c^0||\oln T_c^{*0}$ \\
 $b$ & $\oln B^0||\oln B^{*0}$ & $B^-||B^{*-}$   & $\oln B_s^0||\oln B_s^{*0}$ &
       $B_c^-||B_c^{*-}$       & \fetabU           & $T_b^-||T_b^{*-}$           \\
 $t$ & $T^+||T^{*+}$           & $T^0||T^{*0}$   & $T_s^+||T_s^{*+}$           &
       $T_c^0||T_c^{*0}$       & $T_b^+||T_b^{*+}$ & \frame{$\eta_t||\theta$}    \\
\end{tabular}
\end{table}

\begin{table}
\caption{ The present spectra of vector bottom strange mesons
$B_s^{*}(nS)$ compared with available data from Hagiwara et
al.~\protect\cite{PDG02} and predictions of Godfrey et
al.~\protect\cite{GOD85}. Note that the data for $B_s^{*}(1S)$ is
not confirmed~\protect\cite{GOD85}. Masses are in MeV. }
\label{table:Bs}
\begin{tabular}{lccc}
 $B_s^{*}$  & Ours   & Data~\protect\cite{PDG02} &
  Godfrey et al.~\protect \cite{GOD85} \\
 \hline
 $1^3S_1$   & 5342.2 & 5416.6 & 5450  \\
 $2^3S_1$   & 6123.8 & ---    & 6010  \\
 $3^3S_1$   & 6816.4 & ---    & ---   \\
 $4^3S_1$   & 7444.7 & ---    & ---   \\
 $5^3S_1$   & 8024.2 & ---    & ---   \\
 $6^3S_1$   & 8564.6 & ---    & ---   \\
\end{tabular}
\end{table}

\begin{table}
\caption{
The present spectra of vector bottom charmed mesons $B_c^{*}(nS)$
compared with predictions of Godfrey et al.~\protect\cite{GOD85}
(no data available). Masses are in MeV.
}
\label{table:Bc}
\begin{tabular}{lcc}
 $B_c^{*}$  & Ours   & Godfrey et al.~\protect \cite{GOD85} \\
 \hline
 $1^3S_1$   & 6345.8 & 6340  \\
 $2^3S_1$   & 6830.4 & 6890  \\
 $3^3S_1$   & 7282.9 & ---   \\
 $4^3S_1$   & 7709.0 & ---   \\
 $5^3S_1$   & 8112.6 & ---   \\
 $6^3S_1$   & 8497.1 & ---   \\
\end{tabular}
\end{table}

\begin{table}
\caption{
The present spectra of vector $t$-quark mesons compared with
predictions (in parenthesis) of Godfrey et
al.~\protect\cite{GOD85}. Results with $m_t=35$
GeV~\protect\cite{GOD85} and $m_t$ = 175 GeV~\protect\cite{PDG02}
are given in the first and the third row of each sector. The
masses are in GeV. The spectra of $T^{*0}$ and $T^{*\pm}$ are the
same since $m_u=m_d$ is assumed.
}
\label{table:top}
\begin{tabular}{lcccccc}
 Mesons        & $M(1^3S_1)$ & $M(2^3S_1)$ & $M(3^3S_1)$ &
                 $M(4^3S_1)$ & $M(5^3S_1)$ & $M(6^3S_1)$ \\
 \hline
 $T^{*}$       & 35.24   & 36.27 & 37.27 & 38.24 & 39.19 & 40.12 \\
               & (35.39) &       &       &       &       &       \\
               & 175.24  & 176.27& 177.30& 178.33& 179.34& 180.36\\
 \hline
 $T_s^{*}$     & 35.25   & 36.03 & 36.79 & 37.53 & 38.26 & 38.97 \\
               & (35.46) &       &       &       &       &       \\
               & 175.25  & 176.02& 176.79& 177.56& 178.32& 179.08\\
 \hline
 $T_c^{*}$     & 36.25   & 36.67 & 37.08 & 37.49 & 37.89 & 38.29 \\
               & (36.30) &       &       &       &       &       \\
               & 176.25  & 176.65& 177.06& 177.46& 177.87& 178.27\\
 \hline
 $T_b^{*}$     & 39.45   & 39.71 & 39.96 & 40.21 & 40.47 & 40.72 \\
               & (39.31) &       &       &       &       &       \\
               & 179.44  & 179.68& 179.92& 180.16& 180.40& 180.64\\
 \hline
 $\theta$      & 69.29   & 69.42 & 69.54 & 69.67 & 69.80 & 69.93 \\
               & (68.70) & (69.39) & (69.71) & (69.93) & (70.10) & (70.40) \\
               & 349.24  & 349.29& 349.35& 349.41& 349.46& 349.52\\
\end{tabular}
\end{table}

\begin{figure}
\postscript{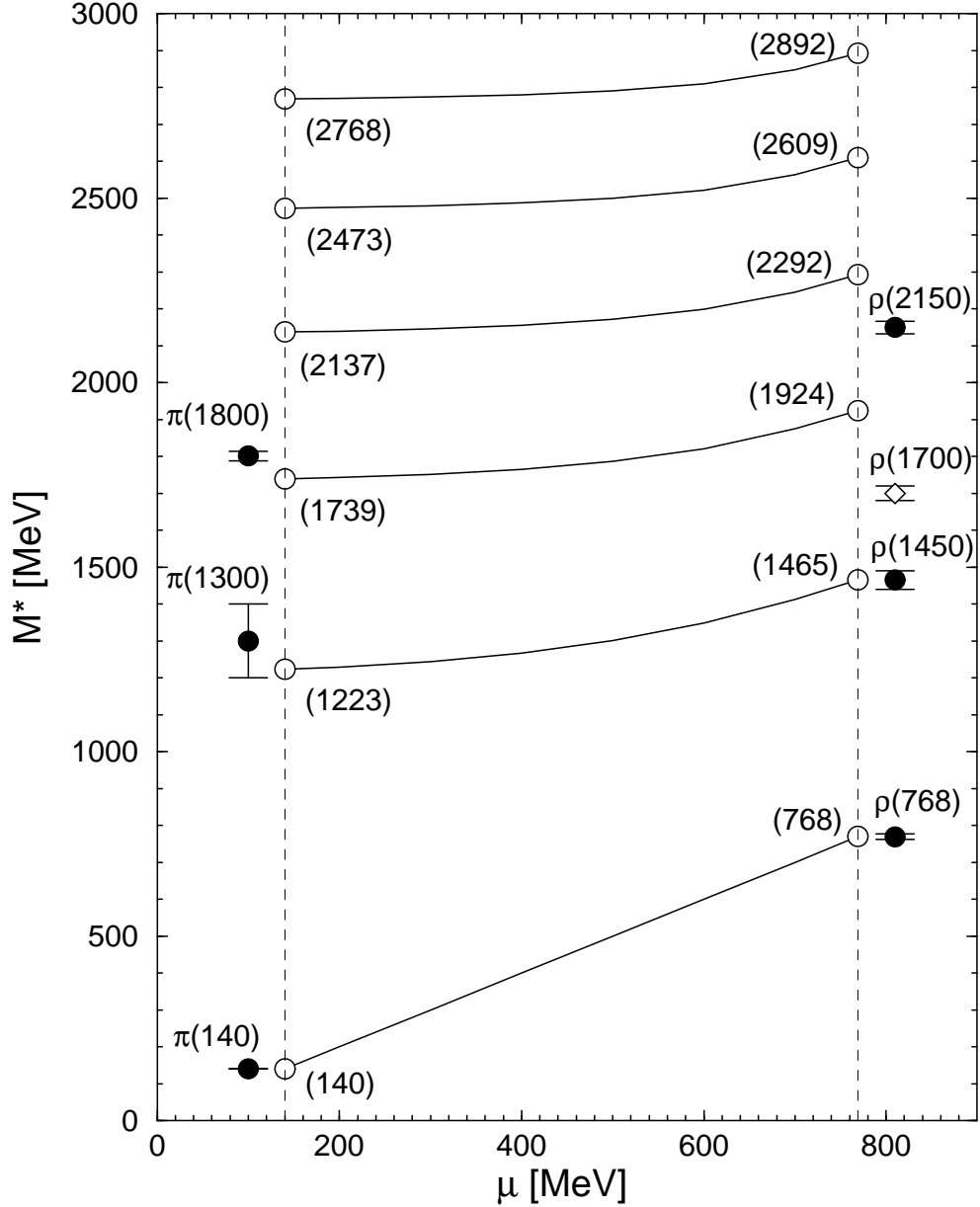}{.8}
\caption{
Mass of the excited $q\oln q$ states ($M^*$) as a function of the
mass $(\mu)$ of the pseudoscalar meson ground state for $I=1$. The
data is taken from Hagiwara et al.~\protect\cite{PDG02} and is
shown by solid circles on the left and right of the figure for
$^1S_0$ and $^3S_1$ mesons, respectively. $\rho(1700)$ (labelled
with empty diamond) might be $D$-wave
dominant~\protect\cite{ANI00,RIK00}. Calculated $^1S_0$ and
$^3S_1$ spectra are given in parentheses within the dashed lines.
}
\label{figure:pi}
\end{figure}

\begin{figure}
\postscript{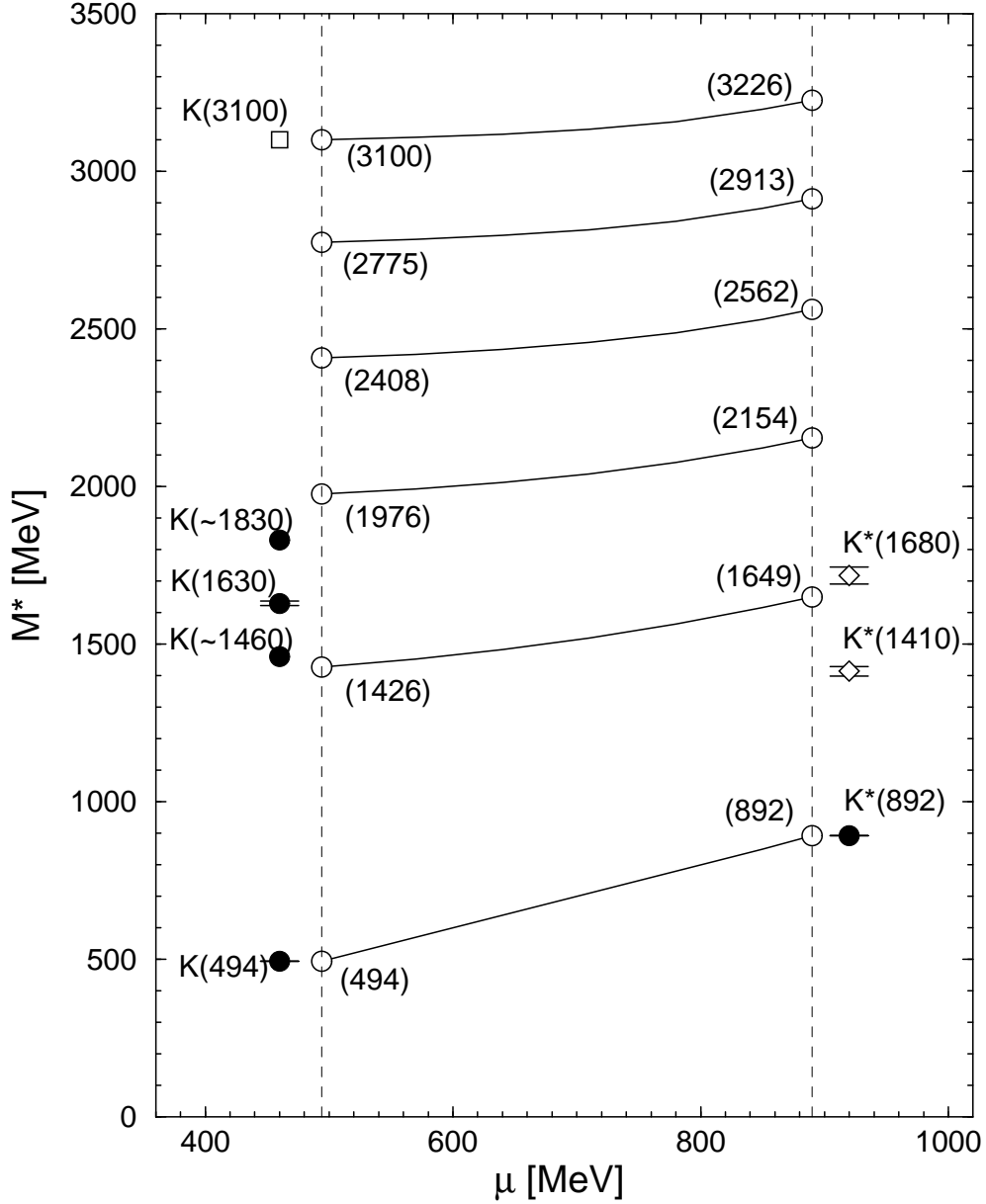}{.8}
\caption{
Mass of the excited $q\oln q$ states ($M^*$) as a function of the
mass $(\mu)$ of the pseudoscalar meson ground state for $I=1/2$ of
the strange mesons. The data is taken from Hagiwara et
al.~\protect\cite{PDG02} and is shown by solid circles on the left
and right of the figure for $^1S_0$ and $^3S_1$ mesons,
respectively. Errors are not available for $K(1460)$ and
$K(1830)$. There is ambiguity about the radial quantum number of
$K^*(1410)$ and $K^*(1680)$~\protect\cite{PDG02} (labelled with
empty diamond). $K(3100)$ is not confirmed and represented by
empty square. Calculated $^1S_0$ and $^3S_1$ spectra are given in
parentheses within the dashed lines.
}
\label{figure:k}
\end{figure}

\begin{figure}
\postscript{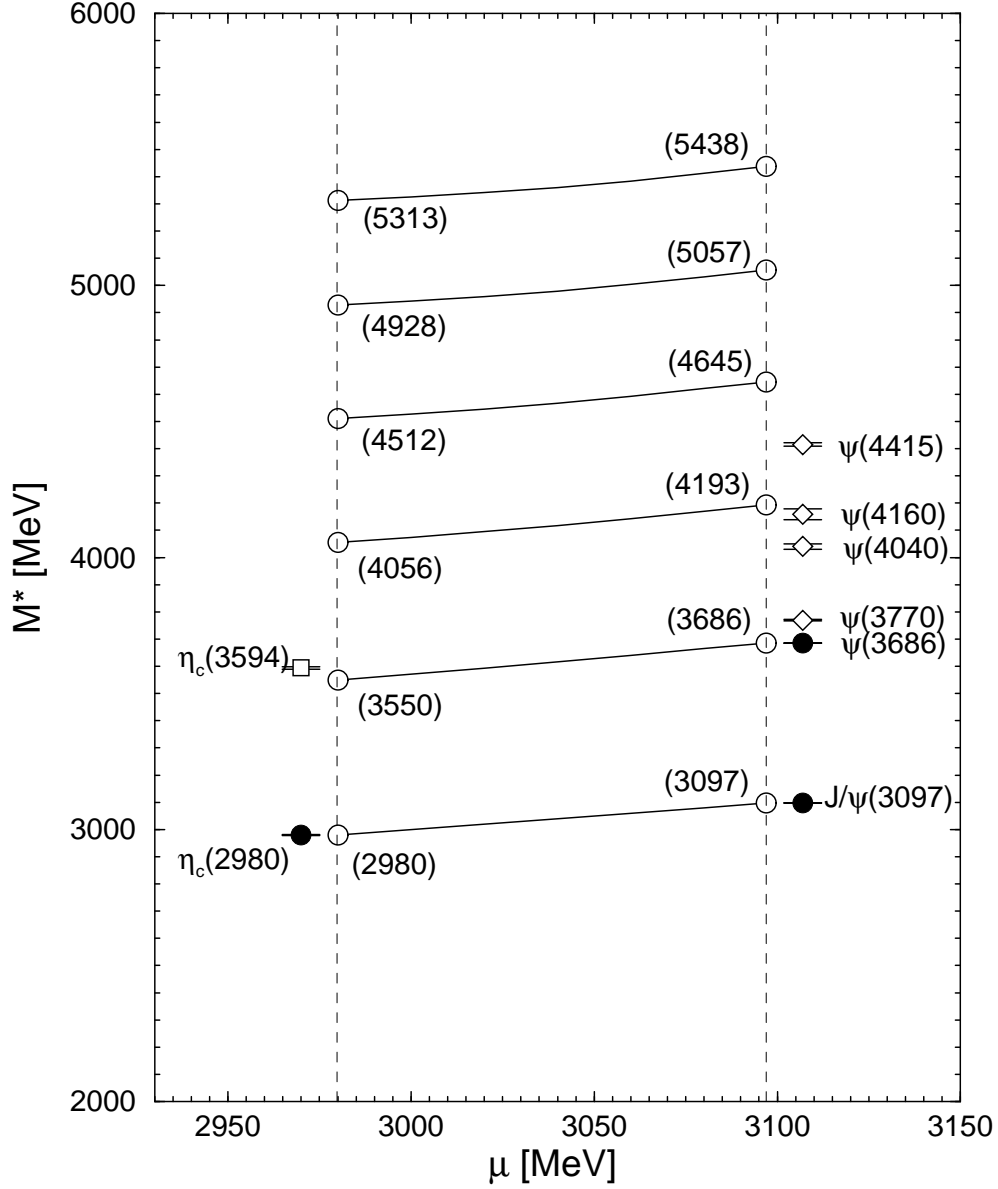}{0.8}
\caption{
Mass of the excited $q\oln q$ states ($M^*$) as a function of the
mass $(\mu)$ of the pseudoscalar meson ground state for $I=0$ of
the charmed mesons. The data is taken from Hagiwara et
al.~\protect\cite{PDG02} and is shown by solid circles on the left
and right of the figure for $^1S_0$ and $^3S_1$ mesons,
respectively. No radial quantum numbers assigned to $\psi(3770)$,
$\psi(4040)$, $\psi(4160)$ and $\psi(4415)$ (labelled with empty
diamond). $\eta_c(3594)$ is not confirmed and represented by
empty square. Calculated $^1S_0$ and $^3S_1$ spectra are given in
parentheses within the dashed lines.
}
\label{figure:etac}
\end{figure}

\begin{figure}
\postscript{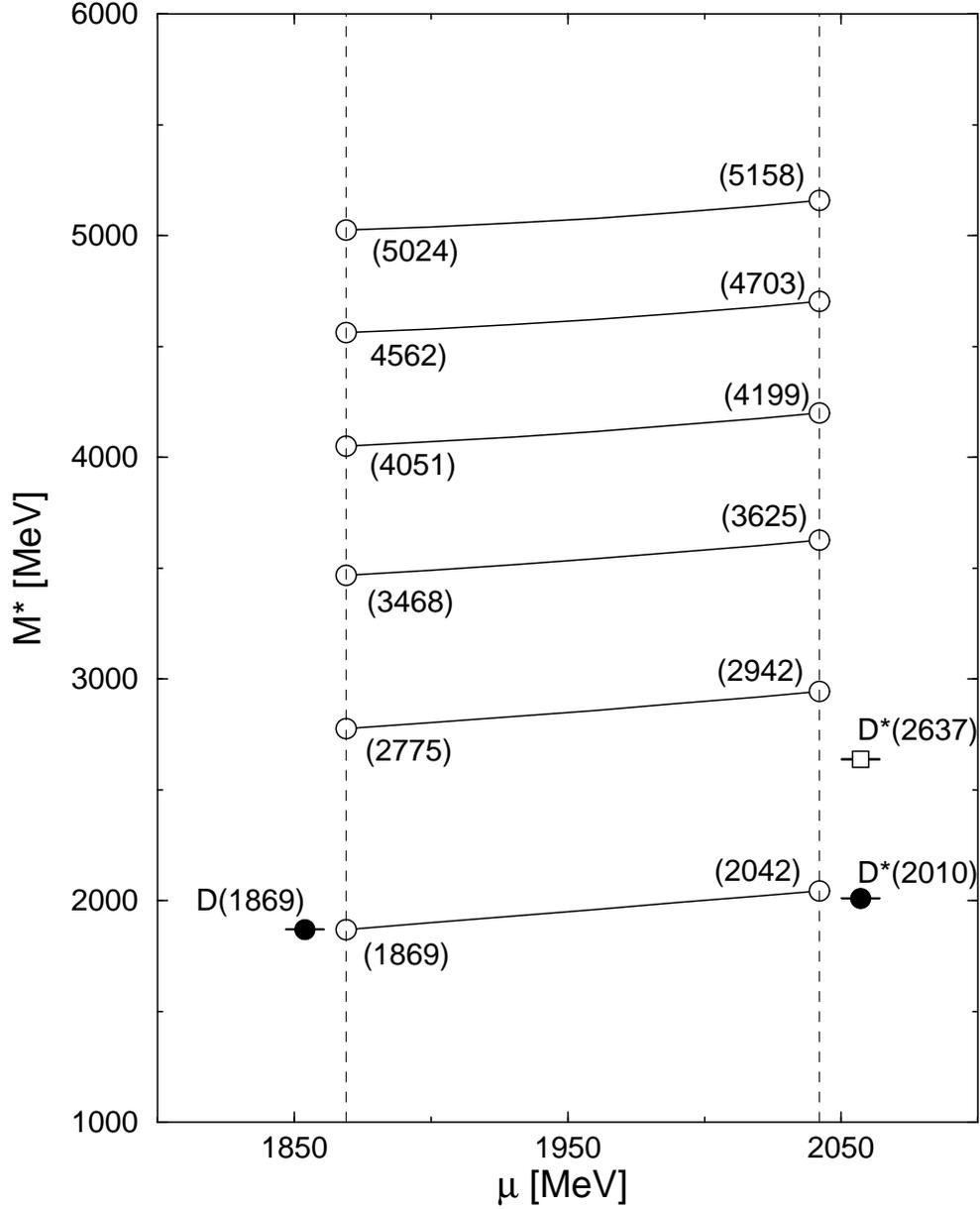}{0.8}
\caption{
Mass of the excited $q\oln q$ states ($M^*$) as a function of the
mass $(\mu)$ of the pseudoscalar meson ground state for $I=1/2$ of
the charmed mesons. The data is taken from Hagiwara et
al.~\protect\cite{PDG02} and is shown by solid circles on the left
and right of the figure for $^1S_0$ and $^3S_1$ mesons,
respectively. $D^*(2637)$ is not confirmed and represented by
empty square. Calculated $^1S_0$ and $^3S_1$ spectra are given in
parentheses within the dashed lines.
}
\label{figure:D}
\end{figure}

\begin{figure}
\postscript{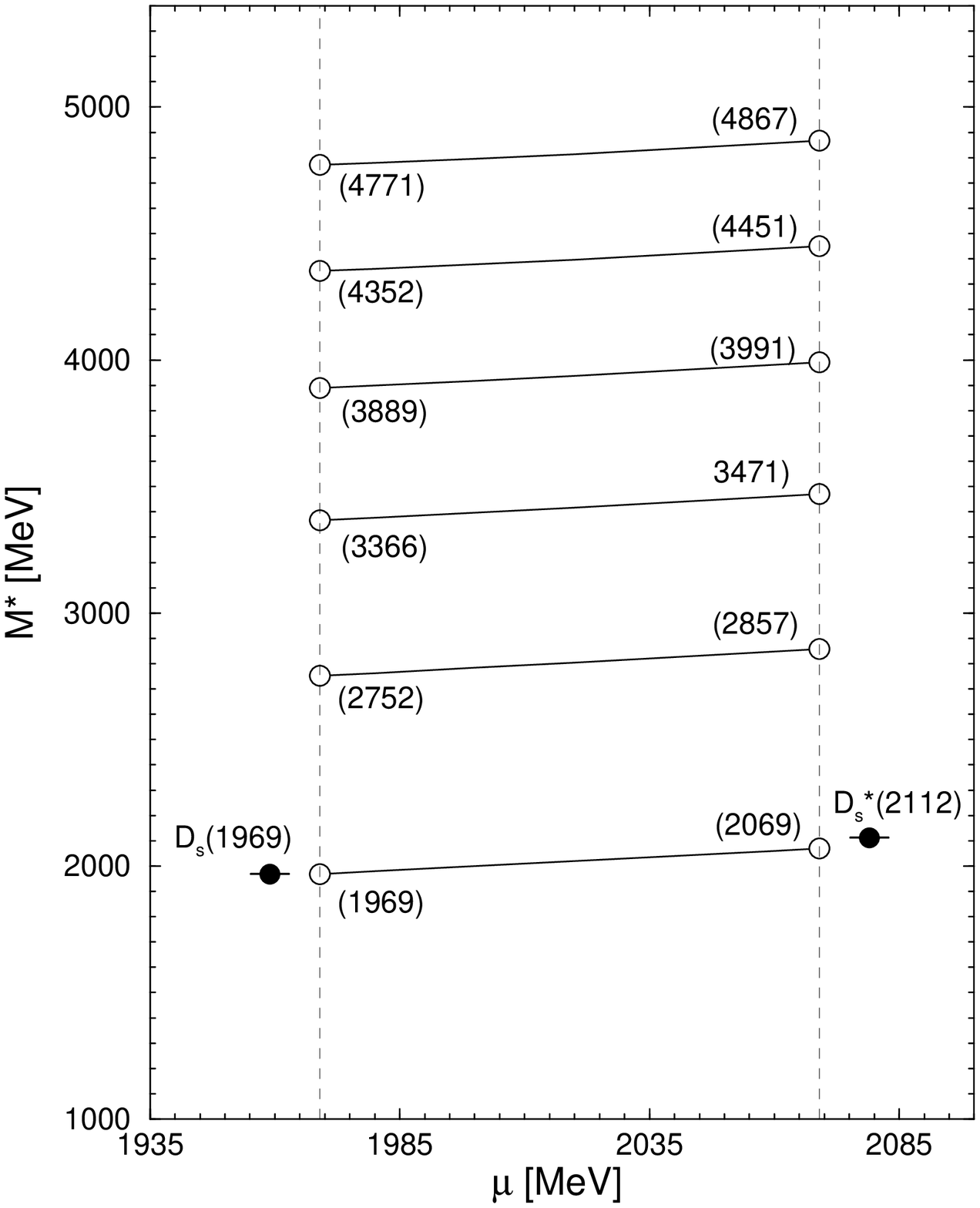}{0.8}
\caption{
Mass of the excited $q\oln q$ states ($M^*$) as a function of the
mass $(\mu)$ of the pseudoscalar meson ground state for $I=0$ of
the charmed strange mesons. The data is taken from Hagiwara et
al.~\protect\cite{PDG02} and is shown by solid circles on the left
and right of the figure for $^1S_0$ and $^3S_1$ mesons,
respectively. Calculated $^1S_0$ and $^3S_1$ spectra are given in
parentheses within the dashed lines.
}
\label{figure:Ds}
\end{figure}

\begin{figure}
\postscript{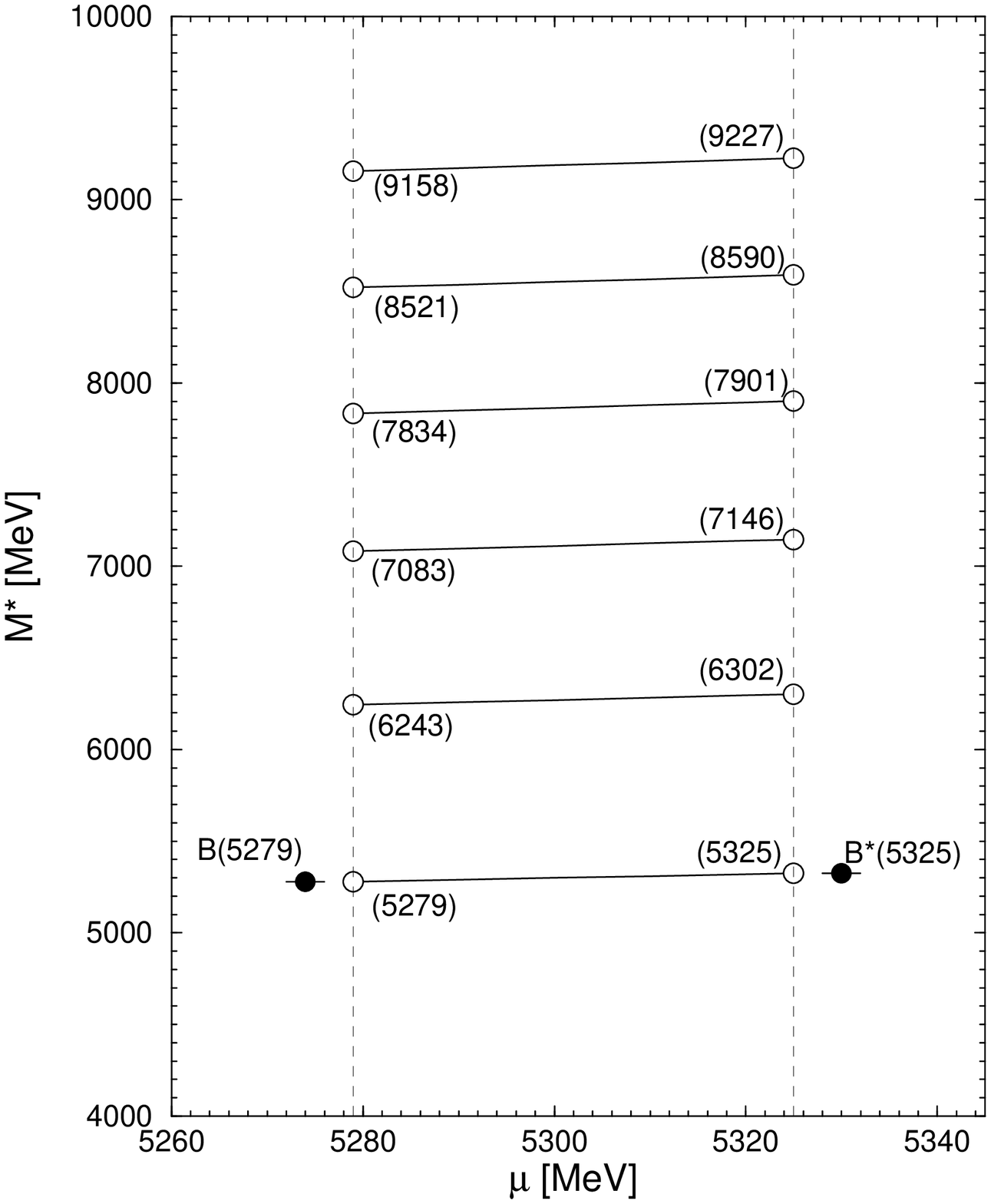}{0.8}
\caption{
Mass of the excited $q\oln q$ states ($M^*$) as a function of the
mass $(\mu)$ of the pseudoscalar meson ground state for $I=1/2$ of
the bottom mesons. The data is taken from Hagiwara et
al.~\protect\cite{PDG02} and is shown by solid circles on the left
and right of the figure for $^1S_0$ and $^3S_1$ mesons,
respectively. Calculated $^1S_0$ and $^3S_1$ spectra are given in
parentheses within the dashed lines.
}
\label{figure:B}
\end{figure}

\begin{figure}
\postscript{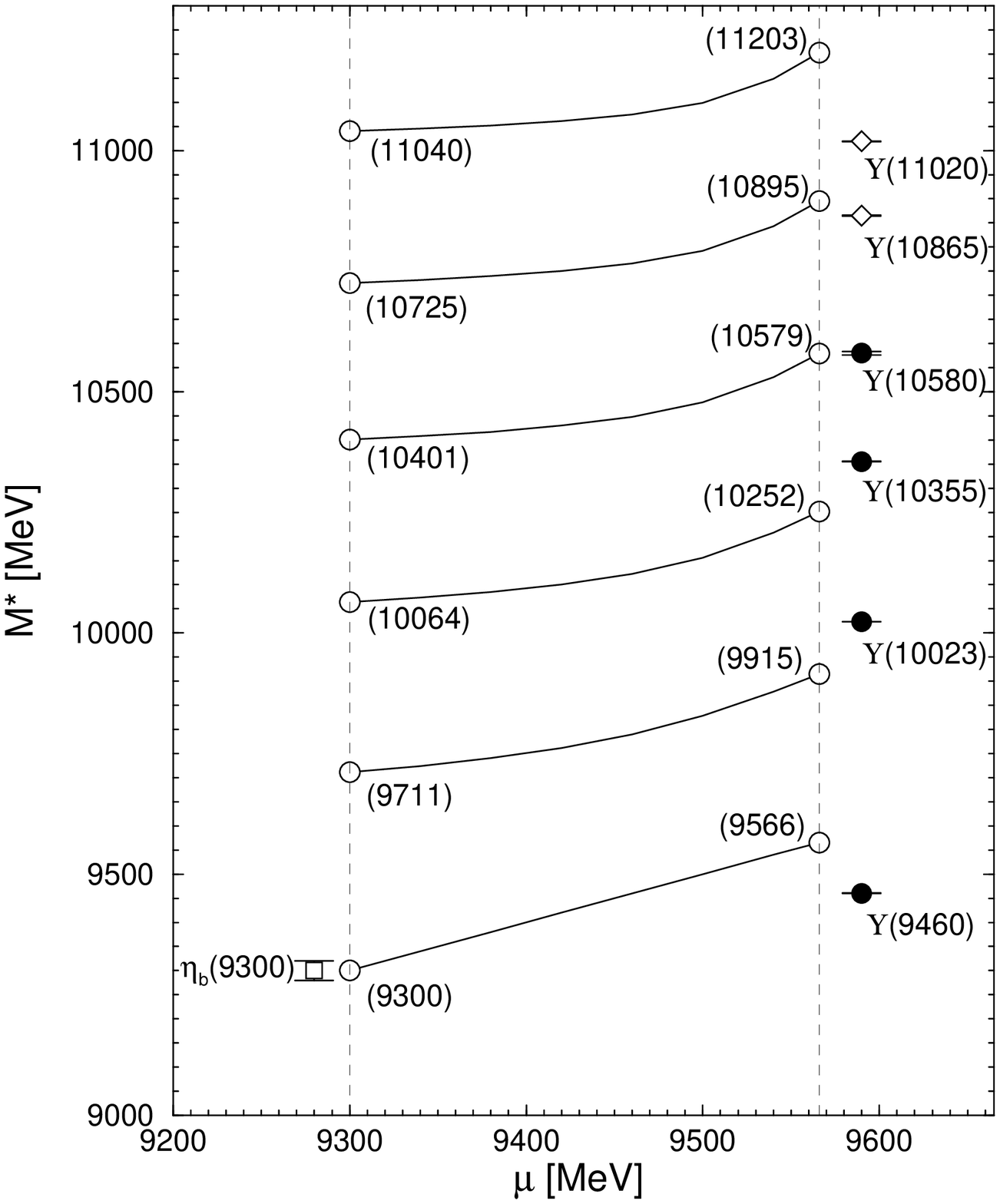}{0.8}
\caption{
Mass of the excited $q\oln q$ states ($M^*$) as a function of the
mass $(\mu)$ of the pseudoscalar meson ground state for $I=0$ of
the bottom mesons. The data is taken from Hagiwara et
al.~\protect\cite{PDG02} and is shown by solid circles on the left
and right of the figure for $^1S_0$ and $^3S_1$ mesons,
respectively. No radial quantum numbers assigned to
$\Upsilon(10865)$, and $\Upsilon(11020)$ (labelled with empty
diamond). $\eta_b(9300)$ is not confirmed and represented by
empty square. Calculated $^1S_0$ and $^3S_1$ spectra are given in
parentheses within the dashed lines.
}
\label{figure:etab}
\end{figure}


\begin{thebibliography}{99}

\bibitem{BRO98} S. J. Brodsky, H.C. Pauli, and S.S. Pinsky,
 Phys. Rep. {\bf 301}, 299 (1998).

\bibitem{PAU00} H.C. Pauli,
 Nucl. Phys. {\bf B} (Proc. Supp.) {\bf 90}, 154 (2000).

\bibitem{FRE01} T. Frederico and H.-C. Pauli,
 Phys. Rev. {\bf D64}, 054007 (2001).

\bibitem{FRE02} T. Frederico, H.-C. Pauli, and S.-G. Zhou,
 Phys. Rev. {\bf D66}, 054007 (2002).


\bibitem{ANI00} A. V. Anisovitch, V. V. Anisovich, and A. V. Sarantsev,
 Phys. Rev. {\bf D62}, 051502(R) (2000);
 V.V. Anisovich, hep-ph/0110326 and references therein.

\bibitem{PDG02} H. Hagiwara et. al.,
 Phys. Rev. {\bf D66}, 010001 (2002).

\bibitem{GOD85} S. Godfrey and N. Isgur,
 Phys. Rev. {\bf D32}, 189 (1985).

\bibitem{FRE00} T. Frederico, A. Delfino, and  L. Tomio,
 Phys. Lett. B {\bf 481}, 143 (2000).

\bibitem{PAU00b} H.C. Pauli,
 Nucl. Phys. {\bf B} (Proc. Supp.) {\bf 90}, 259 (2000).

\bibitem{GOL63} P. Goldhammer,
 Rev. Mod. Phys. {\bf 35}, 40 (1963).

\bibitem{RIK00} R. Ricken, M. Koll, D. Merten, B. Ch. Metsch, and H. R. Petry,
 Eur. Phys. Jour. {\bf A9}, 221 (2000).

\bibitem{BAD02} A.M. Badalian, B.L.G. Bakker and Yu.A. Simonov,
 Phys. Rev. D {\bf 66}, 034026 (2002).

\end{thebibliography}
\end{document}